\documentclass[journal]{IEEEtran}
\usepackage[utf8]{inputenc}

\usepackage{hyperref}
\usepackage{graphicx}
\ifCLASSINFOpdf
\else
\fi
\usepackage{amsfonts}
\usepackage{color}
\usepackage{soul}
\usepackage{subcaption}
\usepackage{tabularx,booktabs}
\usepackage{amsmath,amssymb,amsfonts}
\usepackage{textcomp}
\usepackage{url}
\usepackage{multirow}
\usepackage[table,xcdraw]{xcolor}

\usepackage{boldline}
\usepackage{amssymb}
\usepackage{pifont}

\usepackage[noadjust]{cite}

\begin{document}
	\title{Usage of Network Simulators in \\Machine-Learning-Assisted 5G/6G Networks}
	
	\author{Francesc~Wilhelmi,~Marc~Carrascosa,~Cristina~Cano,~Anders~Jonsson,~Vishnu~Ram,~and~Boris~Bellalta%
		\thanks{Francesc Wilhelmi, Marc Carrascosa, Anders Jonsson, and Boris Bellalta are with Universitat Pompeu Fabra (UPF); Cristina Cano is with Universitat Oberta de Catalunya (UOC); Vishnu Ram is currently working as an independent researcher.}
	}
	
	\maketitle
	
	\begin{abstract}
		Without any doubt, Machine Learning (ML) will be an important driver of future communications due to its foreseen performance when applied to complex problems. However, the application of ML to networking systems raises concerns among network operators and other stakeholders, especially regarding trustworthiness and reliability. In this paper, we devise the role of network simulators for bridging the gap between ML and communications systems. In particular, we present an architectural integration of simulators in ML-aware networks for training, testing, and validating ML models before being applied to the operative network. Moreover, we provide insights on the main challenges resulting from this integration, and then give hints discussing how they can be overcome. Finally, we illustrate the integration of network simulators into ML-assisted communications through a proof-of-concept testbed implementation of a residential Wi-Fi network. 
	\end{abstract}
	
	
	\IEEEpeerreviewmaketitle
	
	\section{Introduction}
	
	Beyond the fifth-generation (5G) of mobile communications systems, namely the sixth generation (6G), Artificial Intelligence (AI), and more precisely Machine Learning (ML), are expected to be pervasively included as part of the network operation, which would entail a huge leap towards optimization, automation, and self-healing. This is possible thanks to the paradigm shift driven by the softwarization of networks -- achieved through Software Defined Networks (SDN) and Network Function Virtualization (NFV) -- which provides the necessary flexibility to empower data-driven approaches.
	
	The integration of ML to communications has started to be considered for the upcoming versions of 5G. This fact is supported by the content already approved by the 3rd Generation Partnership Project (3GPP) for Release 16 (2020) and Release 17 (2021) \cite{3gpp2019study}, which aim to continue improving the efficiency of 5G systems in many domains such as interference mitigation, power consumption, and user mobility, to name a few, and further push for Self-Organizing Networks (SON) with Big Data. Besides, we find of high relevance the contributions made by the International Telecommunication Union (ITU) Focus Group on Machine Learning for 5G and Beyond (FG-ML5G) and the Study Group 13 (SG13), which published specifications on an ML-aware architecture \cite{ITU3172, ITU3174}.
	
	Through the exploitation of the rich amount of available data, ML can overcome the systemic complexity inherited from novel use cases like Vehicle to Everything (V2X) communications, Machine Type Communications (mMTC), and extended reality and high-quality video content delivery. These use cases comprise heterogeneous scenarios with mobility, a huge number of devices, and high-bandwidth and low-latency requirements. In particular, ML may offer substantial performance gains due to the inherent flexibility of automatically learning diverse situations, thus allowing to solve problems related to interference management, improving spatial reuse, or efficient resource allocation. ML mechanisms such as neural networks or Q-learning (to name a few popular methods) have been widely adopted to address communication and security issues in networks (e.g., secure Internet-of-Things access control). The promising benefits of ML-enabled communications systems has been studied in \cite{itu_architecture}. We refer the interested reader to that work and the references therein.
	
	While ML promises significant productivity gains, it also raises serious challenges and concerns. First of all, the successful application of ML depends on the quality of the available training data. An important challenge lies, therefore, in problems with limited or noisy data. Apart from that, dealing with non-stationary data is still an open challenge, which casts doubts on the validity of learned models. A prominent example is that of IEEE 802.11 Wireless Local Area Networks (WLANs). The typical decentralized nature of WLANs (e.g., residential deployments) complicates data collection procedures, and also leads to complex and highly non-stationary environments, where ML may fail to learn in time.
	
	These challenges put into question the worthiness of introducing ML to networking systems. In particular, network operators and other stakeholders may have concerns regarding architectural (e.g., how to train and transfer ML models across a network) and operational aspects (e.g., how to provide trustworthy ML optimizations). While significant efforts have been put towards defining architectural solutions for ML-aware communications \cite{3gpp2019study, ETSI, ITU3172, ITU3174}, the implications of applying ML methods to networks have been barely studied. In this regard, we find an architectural element in \cite{ITU3172} (the ML sandbox) for training and evaluating ML models in a safe environment. The ML sandbox has been defined to mitigate the potential side effects that ML optimizations can have communications systems. Nevertheless, at the date of publishing this paper, the functionalities of ML sandbox remain unspecified.
	
	In this paper, we devise the usage of network simulators as an important way to realize a sandbox in ML-aware communications systems. Network simulators play a crucial role both in academia and industry, and can potentially enable the paradigm shift towards ML-assisted communications. By prototyping complex problems and systems, simulators are key to evaluate new features and technologies, thus boosting innovation. In this regard, we believe that network simulators can contribute to providing reliable and robust ML mechanisms for communications. A plethora of network simulation tools (e.g., ns-3, OMNET++) can be used in the sandbox environment to characterize different types of deployments (e.g., a WLAN in a campus network) according to the operators' use case (e.g., energy saving enabled by deep learning). Simulators are, therefore, envisioned to validate the performance of ML methods in 5G/6G environments (e.g., the 5G New Radio implementation is already available in ns-3). To the best of our knowledge, this is the first work on addressing this emerging issue. The main contributions of this paper are as follows:
	\begin{itemize}
		\item We discuss the main aspects related to the reliability of ML for future communications.
		\item We devise the usage of simulators for training, testing, and evaluating the performance of ML models for communications.
		\item We showcase the potential integration of network simulators within the ITU ML-aware architecture, which is an adaptable and interoperable framework for realizing specific ML-based network functionalities.
		\item We provide insights on practical aspects for the integration of network simulators into ML-assisted communications systems. 
		\item We illustrate the potential advantages of using simulators in ML-assisted networks by applying the outcome of an ML-driven simulation to a residential WLAN testbed.
	\end{itemize}
	
	\begin{figure*}[ht!]
		\centering
		\includegraphics[width=0.75\textwidth]{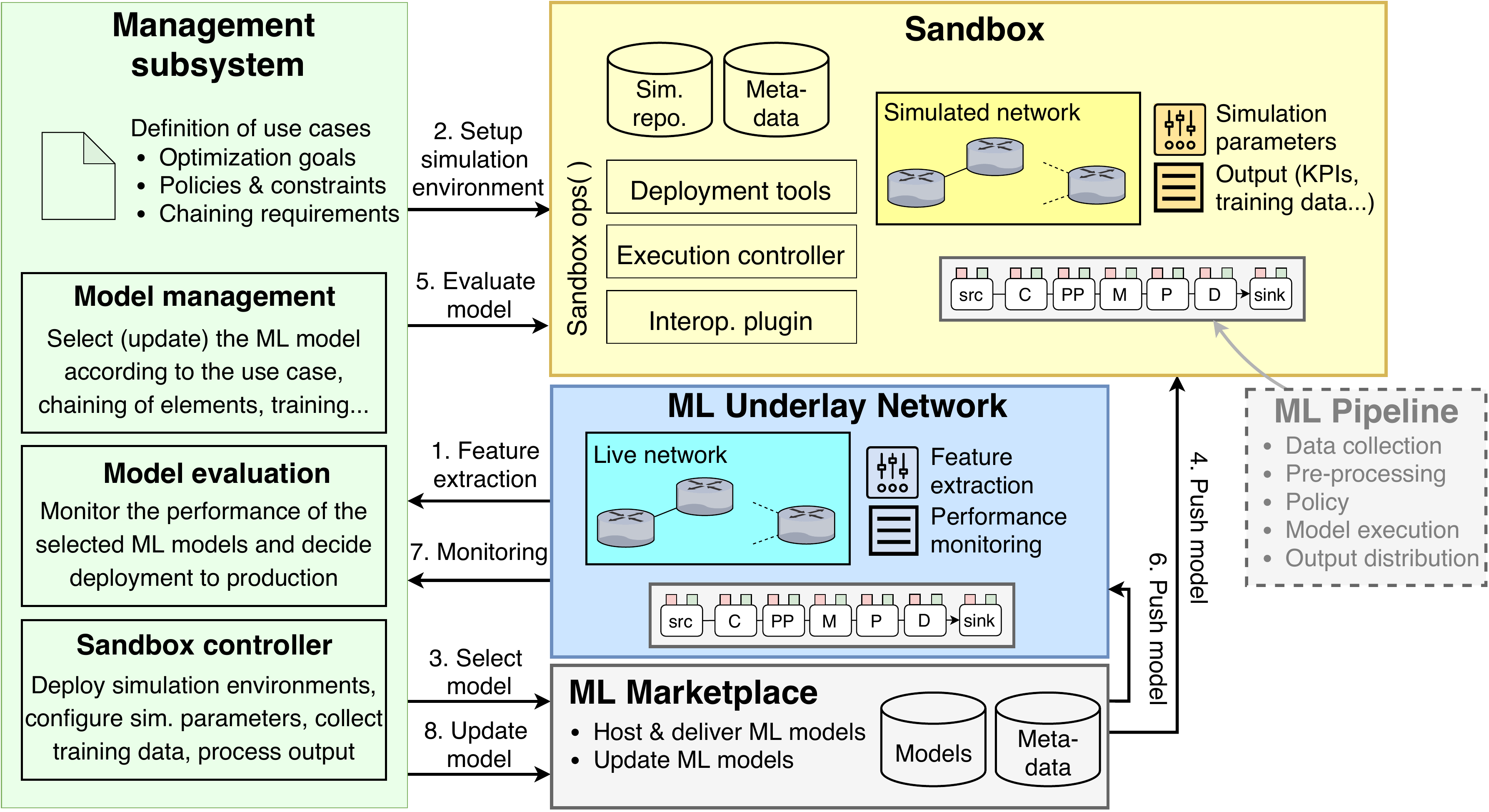}
		\caption{ML-enabling elements and operations of the ITU architecture.}
		\label{fig:example_simulator}
	\end{figure*}
	
	\section{Reliable Artificial Intelligence for Communications}
	
	ML has been proved to offer outstanding results in many fields, but has also raised some concerns in terms of reliability and trustworthiness. The fact is that many ML mechanisms are seen as black boxes due to the non-linearity of their output (e.g., a prediction), especially for problems with high dimensional spaces. For instance, in Deep Learning (DL), the accuracy of a model is typically tied to its complexity in terms of the number of neurons and hidden layers. Although it is possible to get some insights on the behavior of a neural network (e.g., through visualization tools), its inherent complexity hampers interpretability, thus making outputs unpredictable.
	
	In the telecommunications realm, ML has been successfully applied to multiple problems (see, for instance, the surveys in \cite{survey2,survey3,survey4,survey5,survey6} and the references therein). Much of the credit resides in the extraction of useful information from large amounts of data. For instance, the authors of \cite{survey4} show that autonomous Unmanned Aerial Vehicles (UAV) can be empowered by Artificial Neural Networks (ANN). In particular, on-time decisions such as the flying direction can be optimized based on the data collected (e.g., users' location, available resources, or wireless environment). This data, which may come from multiple sources, can be exploited and comprehended by the ANN for the sake of optimization.
	
	Whereas ML can potentially improve the management and operation of networks, the uncertainty associated with the output of ML methods can lead to performance degradation. For instance, an online learning mechanism that is driven by exploration-exploitation may fail to comply with Service Level Agreements (SLAs). The fact is that exploration may lead to experiencing an undesired performance by triggering certain configuration settings. This is a critical aspect to take into consideration since many applications rely on certain minimum requirements to operate, and not meeting them could be even dangerous (for instance, consider networking applications for autonomous driving). As a result, the application of ML can raise concerns and lead to mistrust when applied to networks.
	
	To address the potential lack of confidence generated by ML, we find two main state-of-the-art approaches: (1) \textit{explainable AI} \cite{samek} (understand how ML algorithms behave), and (2) \textit{safe Reinforcement Learning (sRL)} \cite{safe} (provide safe exploration mechanisms). Firstly, explainable AI is based on the interpretation of AI-based decisions, which is useful to devise the impact of potential optimizations and predict misbehavior. However, this field is not mature enough, and the existing techniques are mainly based on visualization, so they are subjective and may lead to misinterpretation. For that reason, explainable AI currently lacks applicability in ML-assisted communications, where the potential understanding of AI mechanisms should be directly translated into specific operations in the network.
	
	Furthermore, sRL aims to minimize the negative effects that unconstrained exploration methods can produce during the learning procedure. This can be achieved either by adding extra information to the exploration mechanism (e.g., external advice) or by applying certain risk-aware criteria (e.g., exploration based on water-filling methods). While sRL is useful to mitigate the randomness of exploration, its application may provide moderate improvements and lead to slow optimization when applied to networks, which may be unacceptable in non-stationary systems. Besides, sRL is restricted only to RL mechanisms, and cannot be generalized to other approaches such as DL.
	
	Given the lack of general mechanisms and procedures for providing trustworthy ML-aware communications, we devise the potential usage of network simulators for training, testing, and evaluating the effect of ML models before being applied to operative networks. In particular, simulators can provide diverse functionalities to enhance the confidence level of future ML-assisted networks: 
	\begin{enumerate}
		\item \textbf{Validate the output of ML models:} a simulator can test and evaluate the output of a certain ML optimization before being applied to a production environment. 
		\item \textbf{Assess the impact of ML models on networks:} simulators can be used to study the effect that a given ML optimization has on the rest of the network. The whole procedure can be simulated together if the simulator includes ML functionalities, which is the case, for instance, of ns-3 and Komondor. To put an example, the authors of \cite{survey6} propose an innovative end-to-end solution to jointly reconstruct the physical layer of a given transmission through DL. In this regard, the usage of simulators can help at devising the effects that the ML-driven approach may produce on the rest of the network layers.	
		\item \textbf{Generate training data:} sometimes, training data extracted from network devices can be sparse, limited, incomplete, or incoherent. To address this, simulators can generate synthetic data, which would broaden the available training data sets. However, assessing the quality of synthetic data sets can be challenging for operators, especially concerning complex problems that cannot be modeled accurately. For that reason, it is important to monitor the effects of applying ML models trained with synthetic data on operative networks.
		\item \textbf{Train ML models:} ML models can also be trained in a simulation environment. As an example, consider the case where online learning is simulated along with the network operation. 
		\item \textbf{Complement ML models:} simulators can also contribute to filling the intersection between model-based and data-driven approaches. The fact is that simulators can act as \textit{experts} to assist the operation of ML algorithms. As an example, random initialization is typically employed for ML methods, which sometimes leads to converging to suboptimal saddle points. By adding additional knowledge from simulations, the learning procedure can be improved.
	\end{enumerate}
	
	\section{Network Simulators to Enable Artificial Intelligence in Communications}	
	In this Section, we describe the architectural aspects of integrating network simulators into ML-assisted communications. Besides, we analyze the key requirements, challenges, and opportunities that emerge from the proposed integration.
	
	\subsection{Architectural Integration}
	Most of the existing simulation platforms have no relation with AI/ML techniques, nor have any integrated module for evaluating and training ML models. Moreover, current simulated network functionalities are typically too specific (e.g., simulate the effect of multiple antennas on the PHY layer performance), and seldom support open interfaces, as a result of being developed by focused academic or industrial organizations. To enable the next generation of ML-based communications systems, it is imperative to design interoperable mechanisms between network simulators and ML mechanisms. For that purpose, we find of high relevance the ITU ML architecture defined in \cite{ITU3172}.	
	
	The ITU ML architecture defines a set of logical components, interfaces, and procedures to realize ML-assisted communications. For a complete overview of the ITU architecture, we refer the interested reader to the work in \cite{itu_architecture}, which proposes a realization for future IEEE 802.11 WLANs, an important part of the 5G/6G ecosystem in unlicensed bands. In particular, the ML-aware architecture is composed of the following elements:
	\begin{itemize}
		\item \textbf{Management subsystem:} this element is responsible for the management and orchestration of the ML operation in a network. The responsibilities of this module range from data collection to ML model deployment and monitoring.
		\item \textbf{ML underlay network:} network at which the ML optimization is applied.
		\item \textbf{ML sandbox:} isolated domain for reproducing the behavior and operation of live networking systems.		
		\item \textbf{ML marketplace:} container of ML models that are applied to operative or simulated underlay networks.
		\item \textbf{ML pipeline:} set of elements that interact with underlay networks to perform the ML optimization. 
	\end{itemize} 
	
	To integrate simulators into the loop of ML-assisted networks, we target interoperability as the principal driver. Interoperability allows building end-to-end ML pipelines in simulated network underlays, thus allowing the integration of network simulators in the ML-aware architecture. The fact is that we find a plethora of proprietary and open-source network simulators (e.g., ns-3, OMNET++, OPNET, NetSim, Komondor) for characterizing multiple types of scenarios, technologies, and network functionalities.\footnote{Besides networking aspects, other specific phenomena can be simulated. For instance, Simulation of Urban MObility (SUMO) and UnderWater simulator (UWsim) simulate vehicular urban mobility and underwater physical effects, respectively, along with OPNET and ns-3 simulators.} Simulators will play an important role in characterizing future 5G/6G use cases (namely, eMBB, uRLLC, and mMTC) because they can cost-effectively reproduce challenging deployments encompassing a massive number of devices with strict network requirements.
	
	To address interoperability, we devise the set of components and procedures to be held at the sandbox that is depicted in Fig. \ref{fig:example_simulator}. In this regard, the management subsystem must deploy, configure, and interact with the simulated functionalities that are required by the ML use case. Notice that, given the diversity of simulation tools (stored and maintained in a repository), an interoperability plugin is required to translate simulator-specific commands into standardized operations. 
	
	The sandbox operations can be enabled by the softwarization of networks in 5G/6G systems. In the first place, this change of paradigm enables all the procedures related to data collection and processing, thus allowing to train and evaluate data-driven models in the sandbox. Besides, virtualized elements and standard interfaces can be reused to build and adapt the network in the simulation domain. For instance, the Open Radio Access Network (O-RAN) project provides an intelligent and virtualized RAN for open hardware. According to that, a RAN controller can be used to handle both the operative network and the equivalent simulation environment.
	
	To illustrate the potential applications of simulators within the ML-aware architecture, Fig. \ref{fig:example_simulator} also provides an example where the output of an ML model is evaluated at the sandbox before being applied to the operative network. The involved procedures are as follows:
	\begin{enumerate}
		\item The management subsystem extracts features from the underlay network.
		\item Based on the characteristics of the underlay network and the ML use case, the simulation environment is prepared through sandbox's deployment tools.
		\item The management subsystem selects the ML model from the marketplace, according to the use-case metadata, the optimization goals, and the available models.
		\item The ML model is pushed into the sandbox and applied to the simulated network.
		\item The ML model is evaluated in the simulator. Evaluation of other ML models may be considered upon unsuccessful results.
		\item Once the evaluation is done, the ML model is pushed into the operative network, where the ML optimization takes place. 
		\item The network performance is monitored, as well as new data is gathered.
		\item The information obtained from monitoring is used to update the ML models and/or metadata in the marketplace.  
	\end{enumerate}
	
	The ML output evaluation procedure allows devising the potential benefits and drawbacks of using a certain ML model in a network. The fact is that ML outputs can sometimes look surprising from the perspective of a network operator, and their effect on the network may be unknown a priori. Unpredictability is further noticeable for complex problems from which there is limited knowledge.
	
	\subsection{Practical Integration Aspects}
	
	Integrating simulators into ML-assisted networks entails a set of challenges concerning execution, interoperability, and portability aspects.
	
	First of all, to test, train, and evaluate ML models in simulators, it is important to reproduce the behavior of the target operative network. The main handicap lies in the plethora of existing simulation tools, each one with specific functionalities and execution requirements. In this regard, it is required to handle simulation-related metadata to maintain information such as the supported technologies and network functionalities, the maturity of simulation blocks (e.g., beta release), and the potential number of domains the simulators can span (e.g., from the core to access network). Metadata can, therefore, enable the automated deployment and configuration of network simulators according to the ML use case requirements. For instance, an update of policies should be reflected in the simulation domain, so that operators' requirements can be fulfilled.
	
	To carry out deployment operations, a great disadvantage is that simulators are written in multiple programming languages (e.g., C/C++, Java) and are supported by different specific platforms. In this regard, containerization (e.g., via Docker) can boost portability and allow network operators to deploy simulators flexibly. Apart from that, parallelization is important to determine, for instance, the number of ML pipeline nodes and simulated network functionalities that the simulator can support at any instant. With knowledge on supported capabilities, the simulated functionalities can be adapted to the use case. For instance, short execution and configuration times can serve to empower ML-driven real-time applications. First, we consider the time it takes for the simulator to generate a given output, which may indicate the tractability of simulating large-scale scenarios. Second, fast reconfiguration of network simulators would allow following potential changes on the operative network (e.g., user demands, available resources, policies, etc.).
	
	Concerning pluggable ML functionalities, built-in ML modules can boost the procedures for simulating the behavior of ML mechanisms or training ML models in a simulator. A few existing simulators support ML functionalities. Two examples are the framework connecting ns-3 with OpenAI Gym \cite{gawlowicz2019ns}, and the agent-based implementation in Komondor. The ns3-gym Gym framework exemplifies the interconnection of network simulators with ML platforms, and opens the door to new integrations with other well-known ML libraries such as TensorFlow or Acumos. In ns3-gym, the synchronization between the simulator and the ML components is achieved through a discretized mapping of the simulation's information (e.g., channel status, nodes number, traffic demands) with the states, actions, and rewards to be used by an ML algorithm. The whole procedure is carried out in execution time.
	
	When it comes to interoperability, an important aspect lies in the degree of flexibility of simulators for interacting with the components of the ML-aware architecture. Interoperability is, therefore, meant to enable a seamless integration of intelligent network functionalities in the communication network. For that, it is imperative that the simulated network functionalities are managed using the same operation and maintenance mechanisms as for the network functionalities in the ML underlay. In this regard, the interoperability plugin is crucial to handle the different simulated networks in execution time, thus allowing for standardized functionalities such as \textit{start} or \textit{stop}. In particular, the features that may facilitate the interoperability of out-of-the-box simulators are the support for Command-Line Interface (CLI) execution mode, the level of monitoring supported (real-time, batch, model-based, etc.), automation of data collection, and in applying the ML output in the simulator (e.g., reading from log files vs. API-based interface with ML functions).
	
	\subsection{Accuracy of Network Simulators}
	The degree of reliability of a network simulator depends on its accuracy on reproducing the actual phenomena. In other words, simulations must be as close as possible to reality. This topic was previously addressed in \cite{accuracy_manet}, where the authors defended that simulators do not really fit the actual behavior of networks, based on experimental results in a MANETs testbed. Nevertheless, it was also shown that simulation results can serve as a good upper-bound for testbed setups.
	
	In general, network simulators accurately reproduce the behavior of protocols in higher levels of the TCP/IP stack. However, they can fail at characterizing complex physical phenomena such as radio propagation, antenna radiation, or energy consumption. As a result, network simulators typically provide accurate qualitative performance results, and help to predict the behavior of real networks under certain circumstances. In contrast, some results may lack quantitative precision, thus deviating from the exact performance that would be experienced in real networking systems. Alternatively, hybrid approaches can be employed for simulating certain layers (e.g., MAC) while taking advantage of the actual interactions that occur in real implementations. Unfortunately, and to the best of our knowledge, there is little literature on this topic. 
	
	\section{A Use-case: Power Control in Residential WLANs}	
	To illustrate the potential of integrating simulators into ML-assisted networks, we provide a testbed implementation of an IEEE 802.11 WLAN that suffers from starvation due to the high sensed interference of a residential environment. To address this problem, a joint ML-based solution is simulated, validated, and then provided to the testbed devices.
	
	\subsection{From Testbed to Simulation Domain}
	The considered testbed implementation comprises two overlapping Basic Service Sets (BSSs) in a residential environment, which are characterized by being highly dense and uncoordinated. The decentralized nature of WLAN deployments in a neighborhood may lead to high interference, which can be extremely variable due to the heterogeneous usage of the network and the complex physical phenomena that can occur. The non-stationarity characteristic of residential environments is one of the most critical aspects to be considered when designing dynamic solutions for improving network performance. Hence, the usage of network simulators can contribute to reducing the performance losses originated by the transitory phases of training (e.g., exploration in online learning).
	
\begin{figure}[ht!]
		\centering
		\includegraphics[width=\columnwidth]{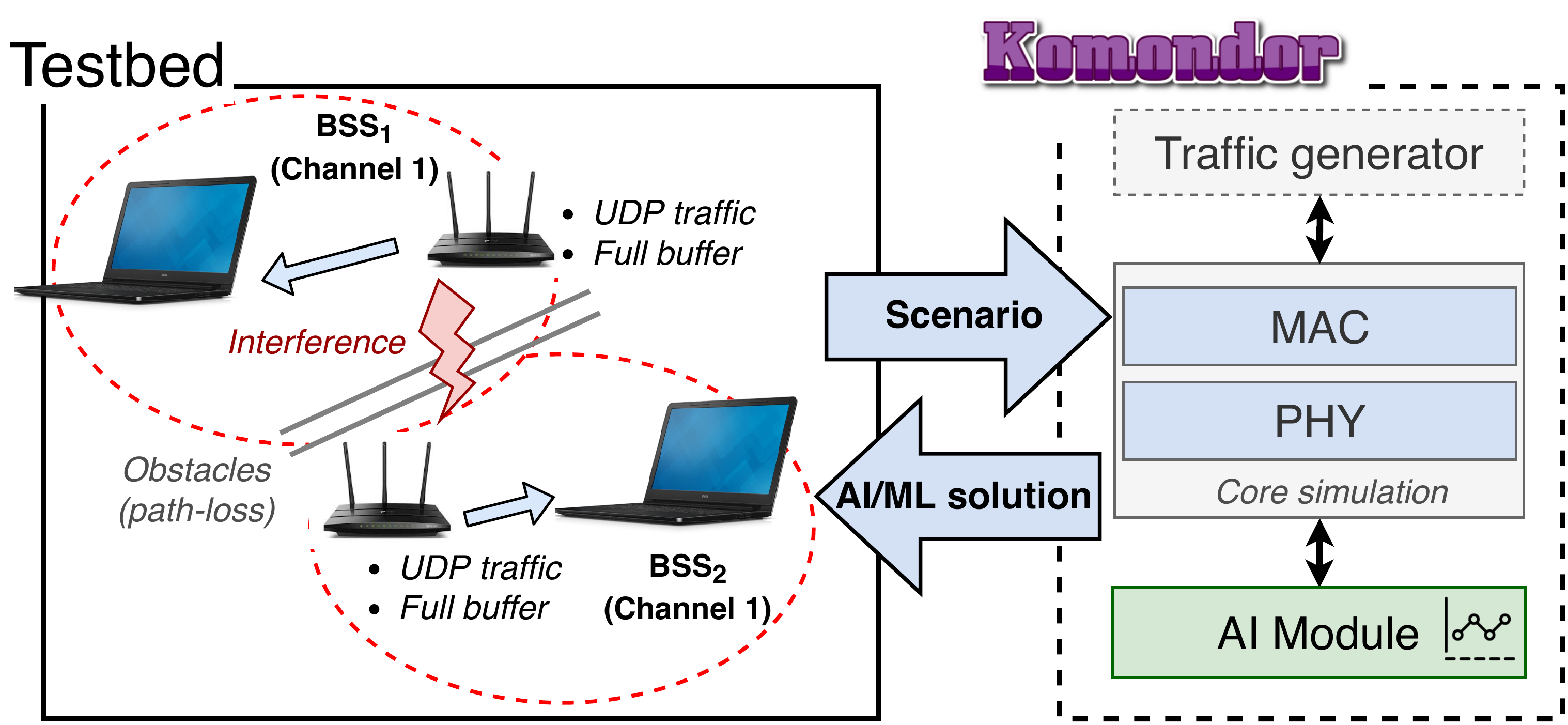}
		\caption{Use case application of the Komondor simulator to apply ML in a testbed WLAN.}
		\label{fig:testbed}
	\end{figure}

	Our proposed testbed-simulator integration is illustrated in Fig. \ref{fig:testbed}, where a tested ML solution is provided to the testbed devices by a simulated version of it. Two identical BSSs are deployed in a high-density residential scenario. However, since they are positioned at different locations, they are subject to different interference conditions, and so experience different performance. The characterization of the WLAN testbed is done with the IEEE 802.11ax-oriented Komondor simulator, which includes the operation of agents for simulating the behavior of ML mechanisms when plugged into wireless nodes.\footnote{All the details of the experimental part and source code are open and available at the following repository: \url{https://github.com/fwilhelmi/usage_of_simulators_in_future_networks}, accessed on May 15, 2020.}
	
	Through the procedures that have been previously illustrated in Fig. \ref{fig:example_simulator}, the testbed scenario is first characterized in the simulator by gathering parameters such as the location of nodes, path-loss effects, or the traffic load. As an example of the characterization of the testbed in the simulator, consider the path-loss model selected, which is chosen based on the degree of similarity with respect to testbed measurements. After preparing the simulation environment, the ML model is applied in the simulator for the sake of improving a certain performance metric. Finally, the optimized ML-based configuration is passed and applied to the real devices whose performance is expected to be enhanced.
	
	\subsection{Machine-Learning-based Transmit Power Control}
	To improve the performance of the target WLAN, we simulate a Multi-Armed Bandits (MABs) mechanism for Transmit Power Control (TPC), as previously done in \cite{wilhelmi}. We take an online learning approach to address the complexity of spatial interactions in WLANs, where the effect of tuning the transmit power can be hindered. Accordingly, the MABs framework is useful to reduce the complexity of the problem and effectively improving the performance at a low computational cost. 
	
	This use case is particularly revealing since the transmit power is a critical parameter to be freely adjusted, and trying several configurations before finding the best performance may lead to unpredictable effects during the transitory regime. Moreover, commercial equipment typically offers a high delay when changing the transmit power or other parameters such as the primary channel. As a result, network simulators can play a crucial role in palliating the negative impact that exploration can have in communications.
	
\begin{figure}[ht!]
		\centering
		\includegraphics[width=0.8\columnwidth]{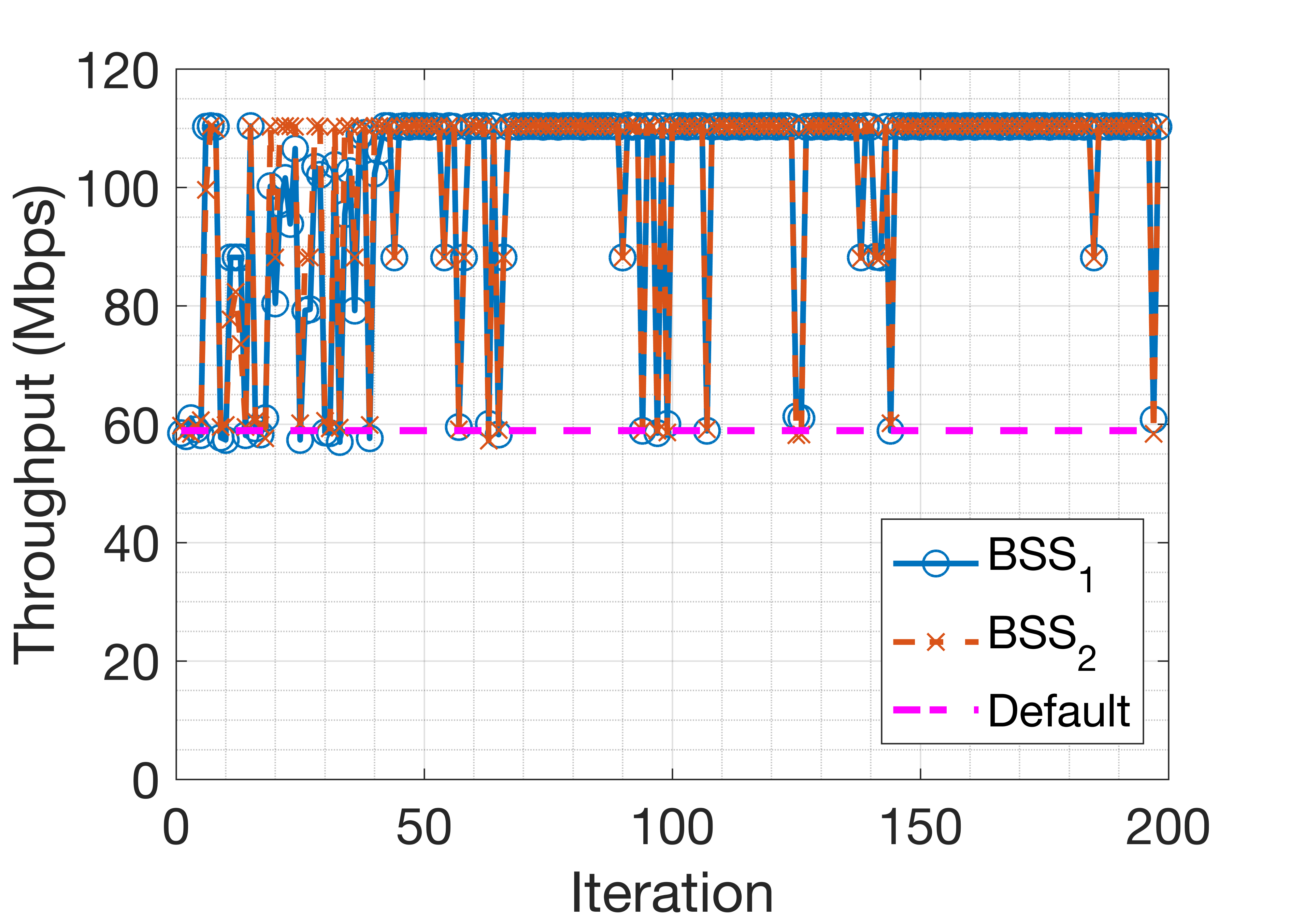}
		\caption{Simulated throughput evolution after applying MABs for tuning the transmit power in an Overlapping Basic Service Set (OBSS). Each learning iteration corresponds to 5 seconds in the simulation.}
		\label{fig:results_komondor}
	\end{figure}

	Figure \ref{fig:results_komondor} illustrates the temporal throughput obtained by each BSS when simulating the MAB method for tuning the transmit power. Also, the performance that would be obtained by both BSSs when using the default configuration is illustrated. As shown, both BSSs experience an unstable transitory regime before reaching a stable state whereby performance is improved. Among the set of input transmit power levels, the most popular one to be used by both BSSs is 7 dBm, which, based on simulation results, is expected to improve the average throughput by 88.48 percent.
	
\begin{figure}[ht!]
		\centering
		\includegraphics[width=0.9\columnwidth]{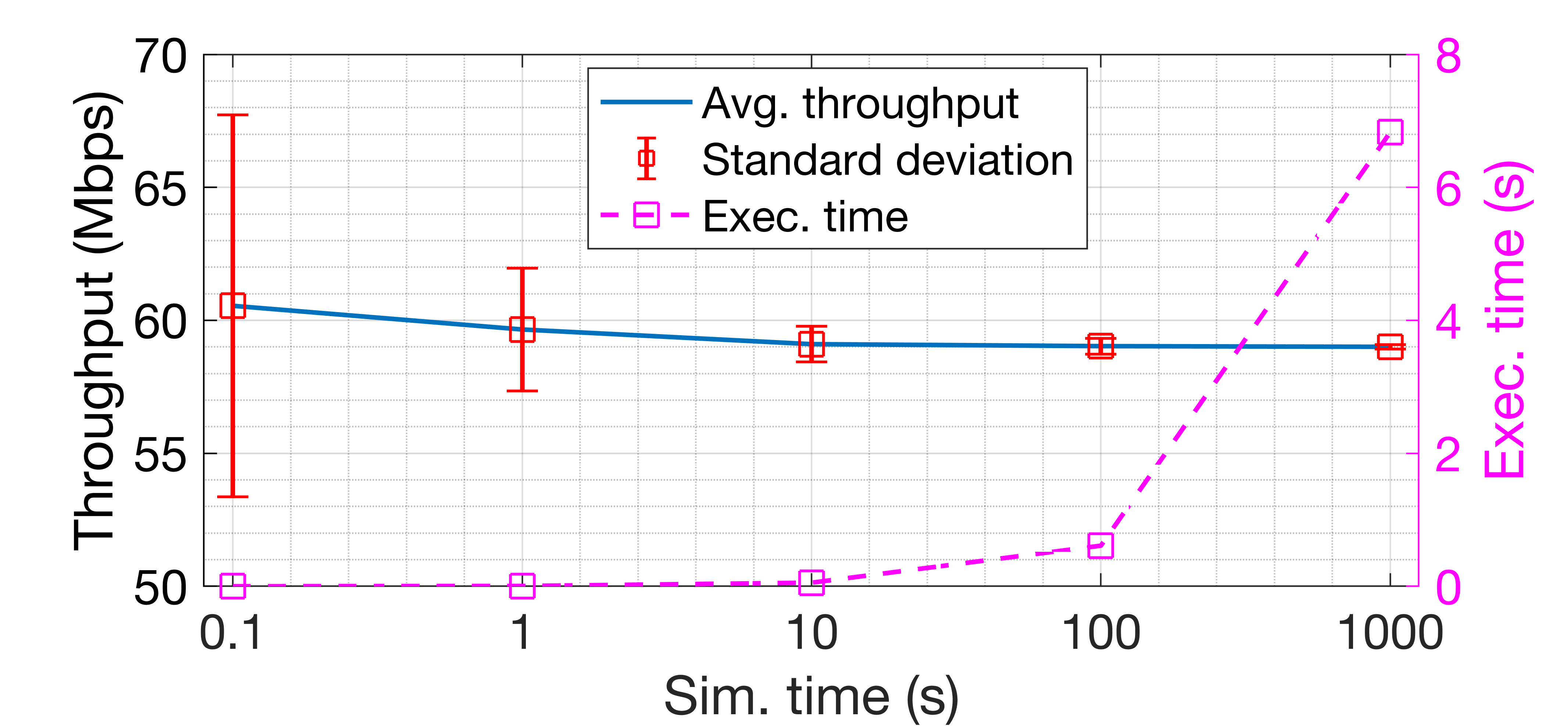}
		\caption{Execution time versus variability of the results in Komondor simulator.}
		\label{fig:test_sim_time_vs_accuracy}
	\end{figure}

	Finally, we provide some insights on the time it takes the simulator to bring up results for the testbed. To include the operation of simulators in future networks (especially for real-time applications), it is very important to find an equilibrium between the stability of the output and the time it takes to generate it. Figure \ref{fig:test_sim_time_vs_accuracy} shows the variability of the simulation results, for different simulation time values. The execution time is also displayed. As observed, the higher the simulation time, the higher the stability is. However, this is paid with execution time, which varies for different network simulators.
	
	\subsection{Testbed Results}
\begin{figure}[ht!!!!]
		\centering
		\includegraphics[width=0.8\columnwidth]{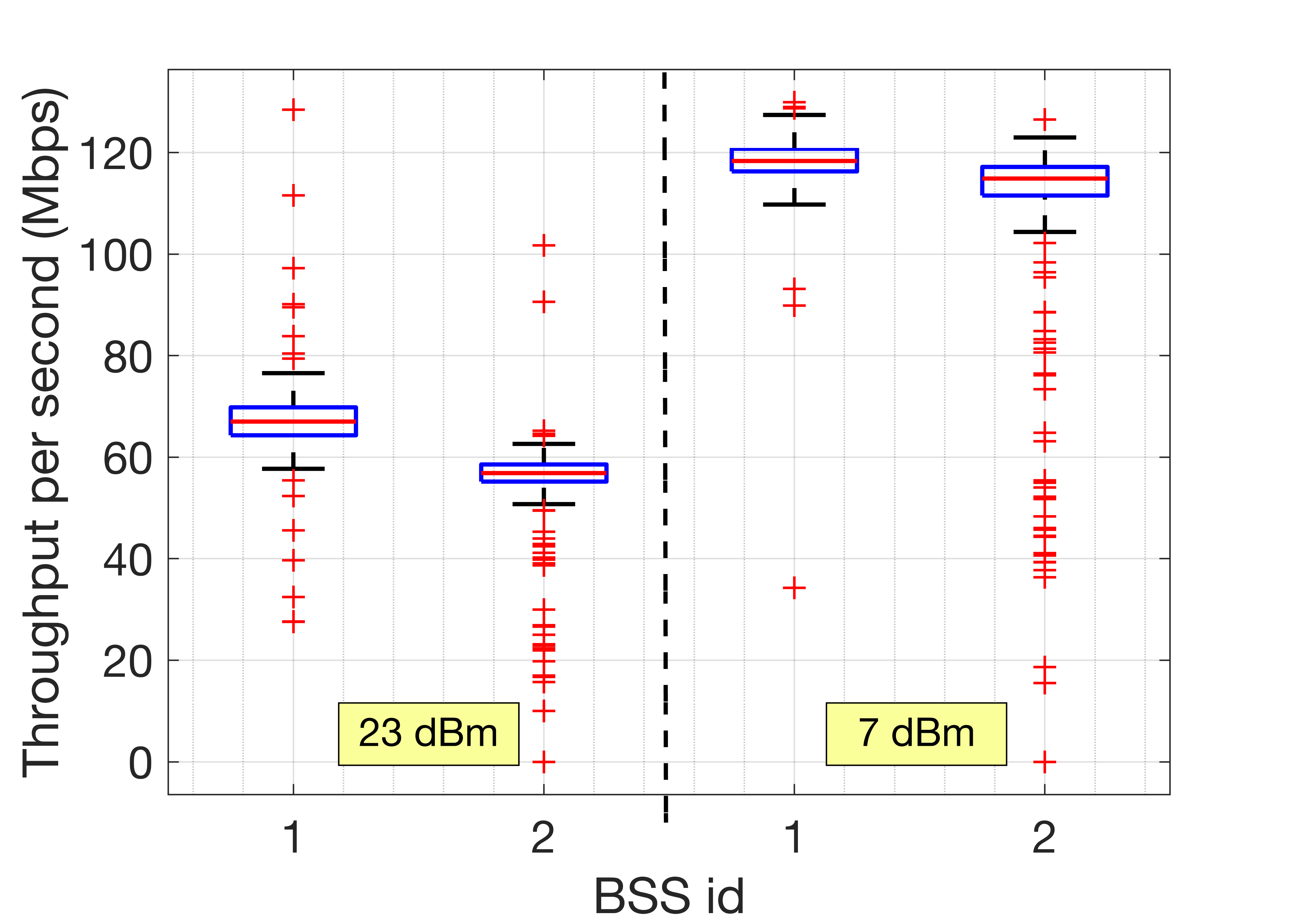}
		\caption{Performance comparison of default (23 dBm) and ML-based (7 dBm) configurations at the testbed WLAN.}
		\label{fig:results}
	\end{figure}

	Now, we show the results of applying the configuration suggested by the simulator on the testbed. Figure \ref{fig:results} compares the performance of applying the ML-based configuration (both BSSs use a transmit power equal to 7 dBm) with that used by default (i.e., 23 dBm).
	
	As shown, both BSSs improve their throughput significantly by using the configuration suggested by the simulator. While BSS$_1$ improves its throughput by 76.16 percent, BSS$_2$ experiences a 93.98 percent improvement. Besides, based on the lower number of observed outliers, we notice a higher stability in terms of throughput variability, especially for BSS$_1$. Note, as well, that BSS$_2$ experiences a higher number of outliers, which are originated by the high channel variability found in the residential environment where the tests were performed.
	
	\section{Concluding Remarks}
	Future communications are expected to evolve towards automated systems enabled by ML. However, the application of ML to networking systems can generate instability and degrade KPIs. To address that, we envision the integration of sandbox environments for ML-assisted networks. In particular, we find network simulators of great utility for training, testing, and evaluating the performance of ML models before being deployed to production environments. In this article, we proposed the utilization of network simulators for future ML-based communications and provided an architectural integration for improving the level of confidence in ML within ML-aware network architectures. Besides, we provided insights into the most prominent challenges resulting from such an integration. Finally, our testbed results in a residential IEEE 802.11 WLAN showed how network simulators allow mitigating the negative effects of directly applying ML in the operative network. 
	
	Network simulators are expected to contribute to filling the gap between AI and communications. Nevertheless, a lot of effort is still needed with regards to the architectural integration of simulators into ML-assisted networks. The most important challenges lie in the definition and implementation of standardized interfaces. Concerning practical implementations, we left as future work the further evaluation of ML models in network simulators. In particular, different ML models and simulation tools can be compared in terms of output accuracy, execution time, or computation needs.
	
	\section*{Acknowledgment}
	This work has been partially supported by grants MDM-2015-0502, WINDMAL PGC2018-099959-B-I00 (MCIU/AEI/FEDER,UE), 2017-SGR-11888, and by SPOTS project (RTI2018-095438-A-I00) funded by the Spanish Ministry of Science, Innovation and Universities.
	
	\ifCLASSOPTIONcaptionsoff
	\newpage
	\fi
	
	\bibliographystyle{IEEEtran}
	\bibliography{bibliography}
	\begin{IEEEbiographynophoto}{Francesc Wilhelmi}
		(francisco.wilhelmi@upf.edu) holds a Ph.D. in Information and Communication Technologies from Universitat Pompeu Fabra (UPF). He is currently a postdoctoral researcher at Centre Tecnològic de Telecomunicacions de Catalunya (CTTC). He has actively contributed to the work of the ITU Focus Group for Machine Learning in Future Networks including 5G (FG ML5G).
	\end{IEEEbiographynophoto}
	\begin{IEEEbiographynophoto}{Marc Carrascosa}
		(marc.carrascosa@upf.edu) obtained his B.Sc. degree in Telematics Engineering (2018) and a M.Sc. in Intelligent and Interactive Systems (2019) from Universitat Pompeu Fabra (UPF). He is currently a PhD student in the Wireless Networking Research Group in the Department of Information and Communication Technologies (DTIC) at UPF. His research interests are related to performance optimization in wireless networks.
	\end{IEEEbiographynophoto}
	\begin{IEEEbiographynophoto}{Cristina Cano}
		(ccanobs@uoc.edu) holds a Ph.D. (2011) in Information, Communication and Audiovisual Media Technologies from Universitat Pompeu Fabra (UPF). She has been a research fellow in the Hamilton Institute of the National University of Ireland, Maynooth (2012-2014), in Trinity College Dublin (2015-2016) and in Inria-Lille in France (first half of 2016). Currently, she is an associate professor at Universitat Oberta de Catalunya (UOC). 
	\end{IEEEbiographynophoto}
	\begin{IEEEbiographynophoto}{Anders Jonsson} (anders.jonsson@upf.edu) is the director of the Artificial Intelligence and Machine Learning group at Universitat Pompeu Fabra (UPF). He received his Ph.D. in computer science in 2005 from the University of Massachusetts Amherst, USA, and has been at UPF ever since.
	\end{IEEEbiographynophoto}
	\begin{IEEEbiographynophoto}{Vishnu Ram}
		(vishnu.n@ieee.org) worked for Motorola/Nokia/Siemens in advanced technologies teams for 21 years. He was a Scientific Advisory Board Associate (SABA) member of Motorola Networks. He has published several drafts in IETF, contributed to ETSI, 3GPP in his role as a senior specialist (Radio Resource Management). He is currently working as an independent researcher.	
	\end{IEEEbiographynophoto}
	\begin{IEEEbiographynophoto}{Boris Bellalta}
		(boris.bellalta@upf.edu) is an Associate Professor in the Department of Information and Communication Technologies (DTIC) at Universitat Pompeu Fabra (UPF). He is the head of the Wireless Networking research group at DTIC/UPF.
	\end{IEEEbiographynophoto}
	\vfill
	
\end{document}